\documentclass{iau}
\usepackage{graphicx}

\title[] 
{How to detect super-massive binary black holes at parsec scales}

\author[Xiang Liu]   
{Xiang Liu }
\affiliation{Xinjiang Astronomical Observatory of CAS, 150 Science 1-Street, Urumqi 830011, China \\[\affilskip]
}

\pubyear{2015}
\volume{312}  
\pagerange{119--120}
 \jname{Star clusters and black holes in
galaxies across cosmic time} \editors{A.C. Editor, B.D. Editor \&
C.E. Editor, eds.}
\begin{document}

\maketitle

\begin{abstract}
It is difficult to find or identifying the binary black holes in
parsec scales, since the dual AGN may be merged quickly. It is
required to explore more possibilities to identifying binary black
holes in parsec scales, we give some discussions, especially with
the VLBI methods.

\keywords{black hole physics: binary -- galaxies: jets -- quasars:
general -- radio continuum}
\end{abstract}



It is possible that super-massive binary black holes live in
active galaxies through merging process. For the huge
gravitational potential of massive black hole and the angular
momentum losses via gas accreting and emission, the binary massive
black holes will be gradually merged to form a larger black hole.
It is not clear that the time scale of the evolution from far
separation of two massive black holes to kpc scale separation, and
the time scale from the kpc scale to pc scale separation.
Statistical studies seem to suggest that the detection rate of
pc-scale binary black holes (BBH) is much less than that of the
kpc-scale dual AGN \cite[(Smith et al. 2010)]{smi10}, this implies
the inspiral of dual black holes may be faster in the pc scale
than in the kpc scale \cite[(Blecha et al. 2013)]{ble13}. It is
difficult to find or identifying the binary black holes in the pc
scale, since the dual AGN would be to merge quickly. It is
required to explore more possibilities to identifying binary black
holes in the pc-scales, we give some discussions in the following.

\subsection{High spectral resolution to reveal the double-peaked
broad lines}

Considering super-massive binary black holes with similar masses
and they both have accretion disks, the two set of similar
emission lines could be detectable, as the double-peaked emission
line AGN found by \cite[Wang et al. (2009)]{wang09}, some of them
were identified to show dual AGN in the kpc scale. When the dual
AGN evolve into pc-scale, their narrow-line regions can be
jointed, but the two broad-line regions may be not jointed yet,
and their broad emission-lines could be double-peaked and
identifiable. One needs higher spectral resolution and sensitivity
to reveal the double-peaked broad lines. Furthermore, it may also
be possible to associate the pc-scale BBH candidate with the
line-of-sight radial velocity shifts of the broad lines if
attributing the shift to the orbital motion of BBH, from
multi-epoch spectral monitoring \cite[(Liu et al. 2014)]{liu14a}.

\subsection{VLBI detection of double twin-jets from binary black
holes at pc-scale}

Very long baseline interferometry (VLBI) at radio can resolve the
radio loud AGN at pc scales, tens of thousands AGN have been
imaged with the high resolution, but the binary black holes in AGN
with the VLBI were not investigated systematically. Assuming radio
loud AGN fraction is 10\%, the fraction for both the dual AGN are
radio loud will be 1\% , this means for the 1\% of dual AGN that
we should be able to detect the double twin-jets from their binary
black holes. Only one AGN has been identified with such double
twin-jets/cores, PKS 0402+379 \cite[(Rodriguez et al.
2006)]{rod06}, with two flat spectrum radio cores of $\sim$7 pc
apart. \cite[Burke-Spolaor (2011)]{bur11} searched for flat
spectrum radio cores from the geodetic VLBI database, found again
only the PKS 0402+379. We have searched for binary black holes
from the astrophysical VLBI databases, found 5 BBH candidates from
$\sim$2000 sources. Our strategy of searching for BBH candidates
is looking for not only flat spectrum double cores, but also
double twin-jets pair or their variant complex \cite[(Liu
2014)]{liu14b}. However, most of the VLBI-imaged jets we
investigated are limited in spatial resolution at centimeter
bands, further investigations to confirm the BBH candidates are
ongoing through high and multi-frequency VLBI observations.

\subsection{VLBI core-offset to be determined as due to the BBH
orbital motion}

\begin{figure}
\centering
    \includegraphics[width=4cm]{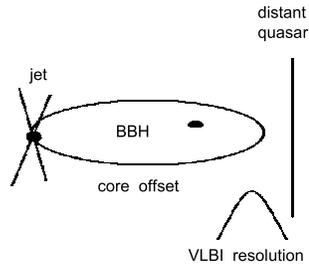}
    \caption{The schematic diagram of the phase-reference VLBI method to detect
    the core-offset of the radio loud core in a binary black hole system.}
     \label{fig1}
  \end{figure}

For binary black hole AGN, with radio loud probability 10\% of
each hole, it is likely that only one BH is jetting and the other
BH is radio quiet in most of dual AGN. In this case, one cannot
detect twin jets from both black holes. The BL Lac object OJ287,
for instance, is believed to host a BBH system for its nearly 12
years periodic optical/radio outbursts that attributing to the
orbital motion of BBH. The VLBI image of OJ287 shows only
one-sided jet, exhibiting large position-angle changes of
inner-jets at high frequencies \cite[(Liu et al. 2012)]{liu12} and
with the core-shifts between different frequencies in VLBI images
as well. The core shifts are mainly caused by the opacity effect
of inner jets. The position-angle changes of inner jets, however,
may be caused by the precession or orbital motion of the BBH
system. Therefore, with multi-epoch VLBI observations at {\it
same} frequency, one would be able to detect the orbital motion of
the radio loud BH in the BBH system if the other is radio quiet,
the so-called `core offset', with the phase-reference VLBI
technique (see Fig.~\ref{fig1}), as referenced to a distant
high-redshift quasar (which has no detectable core-offset). This
needs a long time to measure the core offset caused by the
pc-scale orbital motion of BBH, but it is plausible to detect for
the binary black holes at low redshift.

\begin{thebibliography}{}

\bibitem[Blecha et al. (2013)]{ble13}
{Blecha, L., Loeb, A., \& Narayan, R.} 2013, \textit{MNRAS}, 429,
2594

\bibitem[Burke-Spolaor (2011)]{bur11}
{Burke-Spolaor, S.} 2011, \textit{MNRAS}, 410, 2113

\bibitem[Liu et al. (2014)]{liu14a}
{Liu, X., Shen, Y., Bian, F.Y., et al.} 2014, \textit{ApJ}, 789,
140

\bibitem[Liu (2014)]{liu14b}
{Liu, X.} 2014, \textit{J. Astrophys. Astr.}, Volume 35,  Number 3
(arXiv1304.1955)

\bibitem[Liu et al. (2012)]{liu12}
{Liu, X., Mi, L.-G., Liu, B.-R., Li, Q.-W.} 2012, \textit{Ap\&SS},
342, 465

\bibitem[Rodriguez et al. (2006)]{rod06}
{Rodriguez, C., Taylor, G.B., Zavala, R.T., et al.} 2006,
\textit{ApJ}, 646, 49

\bibitem[Smith et al. (2010)]{smi10}
{Smith, K.L., et al.} 2010, \textit{ApJ}, 716, 866

\bibitem[Wang et al. (2009)]{wang09}
{Wang, J.-M., et al.} 2009, \textit{ApJ} (Letters), 705, L76


\end{thebibliography}
\end{document}